\documentclass[%
 reprint,
 amsmath,amssymb,
 aps,
 natbib=false,
 authoraddress
]{revtex4-2}
\usepackage{graphicx}
\usepackage{dcolumn}
\usepackage{bm}
\usepackage[utf8]{inputenc} 
\usepackage[T1]{fontenc}    
\usepackage{textalpha}
\usepackage[ngerman,english]{babel}
\usepackage[titletoc,title]{appendix}

\usepackage{amsfonts}  
\usepackage{amsmath}   
\usepackage{amssymb}   
\usepackage{fancyhdr}
\usepackage{parskip}
\usepackage{subcaption}
\usepackage{multirow}
\usepackage{mathtools}
\usepackage{leftindex}
\usepackage{xcolor}

\newcommand{\re}[1]{\mathrm{Re}\left[#1\right]}

\newcommand{\pwisein}{\left\{ \begin{array}{ll}}
\newcommand{\pwiseout}{\end{array}\right.}

\def\thuphys{Department of Physics and Center for High Energy Physics, Tsinghua University, Beijing 100084, China} 

\begin{document}

\preprint{APS/123-QED}

\title{Extraction of All Partial-Wave Interaction Parameters via Femtoscopy:\\ The Central-Force Case}

\author{Renjie Zou}
\email{zourj24@mails.tsinghua.edu.cn}
\thanks{equally contributed.}
\affiliation{\thuphys}

\author{Sheng Xiao}
\email{xiaos20@mails.tsinghua.edu.cn}
\thanks{equally contributed.}
\affiliation{\thuphys}

\author{Zhigang Xiao}
\affiliation{\thuphys}


\date{\today}

\begin{abstract}
We present a universal femtoscopic method to extract the source size and central-force interaction parameters for arbitrary partial waves from experimental data, with the interaction parameters defined in the effective-range expansion. The method is based on the Wronskian determinant and is applicable to both the Coulomb and non-Coulomb cases. This approach is numerically verified against simulated neutron-neutron and proton-proton correlation data obtained from the Reid potential.
\end{abstract}

\maketitle



\section{Introduction}
Femtoscopy has recently gained prominence as a probe of low-energy two-body scattering, owing to its sensitivity to the nuclear matter behavior in collisions and its complementary role to conventional beam experiments~\cite{Recent1,Recent2,Recent3}. Conventional beam experiments typically address short-lived or neutral particles, which are hard to form into beam or target material~\cite{specialbeam} but are abundantly produced in ion–ion and proton–proton collisions. The interactions about such particles constitute a subject of active theoretical and experimental interest~\cite{Exotic1,Exotic2,Exotic3,Exotic4,Exotic5}.

The partial-wave method~\cite{partialwavemethod} is central to many theoretical descriptions of two-body interactions in femtoscopy, serving to construct the kernel wave function. Among the available frameworks, the Lednický–Lyuboshitz (LL) model~\cite{LL1,LL2} has been widely applied to femtoscopic data~\cite{LLapp1,LLapp2}. It offers a simple ansatz based on an asymptotic wave function that includes only the s-wave ($l=0$), which makes it computationally efficient for large datasets and has been thoroughly validated in practical applications. However, owing to its restriction to $l=0$, several studies have sought to incorporate contributions from higher partial waves~\cite{HigherWavePrevious1,HigherWavePrevious2}. The first difficulty encountered when attempting to extend the s-wave framework to higher $l$ is that the Riccati-Bessel function of the second kind diverges for $l>0$. The second is the evaluation of correction terms beyond the s-wave. Various strategies have been adopted in the literature, including replacing the asymptotic wave function with a Born-approximation wave function~\cite{HigherWavePrevious2} and introducing an ad hoc cutoff on the divergent functions~\cite{HigherWavePrevious1}.

In experimental domains, femtoscopy has been extensively utilized across a wide range of collision systems and energy regions, with numerous particle-pair combinations investigated~\cite{Exotic1,LLapp2,widerangeofsys1,widerangeofsys2}. In many experiments, signatures of higher partial-wave contributions have been observed with sufficient precision for quantitative extraction of interaction parameters~\cite{HigherWaveExperi1,HigherWaveExperi2,HigherWaveExperi3}. Theoretically, such enhancements are commonly attributed to the presence of a bound state or resonance, and thus carry significant physical implications, highlighting the need of including high partial waves in theoretical calculations.

As part of a systematic study of femtoscopic physics beyond the s-wave, we plan to present a series of articles on the extraction of interaction parameters for the central- and noncentral-force cases. The staged objectives of the series are to investigate the noncentral interaction of the neutron-proton pair and to compare the resulting parameters with the deuteron electric quadrupole moment measured in other experiments~\cite{ElecQuad1,ElecQuad2}.  This paper is the first in the series, since certain results for the central-force case are prerequisites for treating the noncentral interaction.  The noncentral-force case will be addressed in the following-up paper.

The main goal of this paper is to construct a reliable method for extracting central-force interaction parameters for higher partial waves, while maintaining compatibility with the LL model. The framework also extends to the Coulomb case. To this end, we employ the Wronskian determinant~\cite{Math1} to compute the correction terms and counteract divergence arising from the asymptotic wave function. The analytical form of these correction terms is verified numerically. To assess the practical utility of the method, we further generate simulated neutron-neutron correlation functions using the Reid potential, extract the interaction parameters via our procedure and compare the results with reference values obtained directly from the same potential by the variable phase method. 


\section{Methods}

In the theoretical framework~\cite{LL1,AHP,KooninPratt}, the correlation function $C(\textbf{k},\textbf{K})$ for a pair of particles with a relative momentum $\textbf{k}$ and a total momentum $\textbf{K}$ is defined as 
\begin{equation}
\label{CF}
C(\textbf{k},\textbf{K})=\int d^3r \left| \Psi(\textbf{k},\textbf{r}^\star(\textbf{r},\textbf{K})) \right|^2 S(\textbf{r}),
\end{equation}
where $\Psi$ is the time-reversed exact scattering wave function for an incident relative momentum $\textbf{k}$, evaluated at the spatial coordinate $\textbf{r}^\star$ in the pair-rest frame. The function $S(\textbf{r})$ describes the distribution of the source emitting particle pairs, defined in the source-rest frame at the coordinate $\textbf{r}$. The two coordinates are related by a Lorentz transformation parametrized by $\textbf{K}$.

While $S(\textbf{r})$ is usually specified by a preset form with one or more parameters, such as the Gaussian distribution, the determination of $\Psi$ is the primary task in most femtoscopic models. 
For any practical femtoscopic model, the form of $\Psi$ must satisfy several essential requirements. First, the integrand in Eq.~\eqref{CF} should be amenable to efficient numerical integration, which favors a simple expression for $\Psi$. Second, the parameters to be extracted from the fit must have well-defined and distinct physical meanings that are independent of the relative momenta $\textbf{k}$.

The above requirements can be met by replacing $\Psi$ with its asymptotic form, denoted by $\Phi$. The correlation function can then be written and simplified as
\begin{equation}
\label{newCF}
C(\textbf{k},\textbf{K})\approx\int d^3r \left| \Phi \right|^2 S(\textbf{r})-S(\textbf{0})\int d^3r (|\Psi|^2-|\Phi|^2),
\end{equation}
where we have omitted the independent variables $\textbf{k}$ and $\textbf{r}^\star$. The approximation in Eq.~\eqref{newCF} relies on the assumption that the range of the strong interaction is sufficiently short compared with the source size. This assumption is also adopted in other frameworks, such as the LL model~\cite{LL1}, and its systematic uncertainty will be discussed later. The partial wave expansions of $\Psi$ and $\Phi$ read~\cite{LL1,QM1}
\begin{subequations}
\begin{equation}
\label{Psi_partialwave}
\Psi(\textbf{r})=\sum_{l=0}^\infty u_l(r)P_l(\cos{\theta}),
\end{equation}
\begin{equation}
\label{Phi_partialwave}
\Phi(\textbf{r})=\sum_{l=0}^\infty v_l(r)P_l(\cos{\theta}),
\end{equation}
\end{subequations}
where $u_l$ and $v_l$ are the radial wave functions, satisfying $\lim_{r\to \infty}u_l(r)=v_l(r)$. From Eqs.~\eqref{Phi_partialwave} and \eqref{Psi_partialwave}, the correction term on the right-hand side (RHS) of Eq.~\eqref{newCF} can be reduced to
\begin{equation}
\label{u2-v2}
\int d^3r(\left| \Psi \right|^2-\left| \Phi \right|^2)=4\pi\sum_{l=0}^\infty \int_0^\infty dr (u_l^2-v_l^2).
\end{equation}
It is worth mentioning that the first term on the RHS of Eq.~\eqref{newCF} cannot be reduced like Eq.~\eqref{u2-v2}, caused by the violation of spherical symmetry under the Lorentz transformation. Owing to the singular behavior of $\Phi$ and $v_l$ near the origin, the integration interval must exclude the neighborhood of $r=0$, with the cutoff removed only in the limit. Eq.~\eqref{newCF} is thus interpreted as a limiting expression.

The Wronskian determinant offers a shortcut for evaluating the integral on the RHS of Eq.~\eqref{u2-v2} without the need to solve the explicit radial wave functions. In the following, we address the non-Coulomb and the Coulomb cases separately.

\subsection{The non-Coulomb case}

The non-Coulomb limit of Eq.~\eqref{u2-v2} can in principle be obtained from the Coulomb result by taking the Sommerfeld parameter $\eta\to 0$. Nevertheless, we first treat the non-Coulomb case separately to familiarize ourselves with the procedure. The radial Schrödinger equations satisfied by $u_l$ and $v_l$ are written as~\cite{QM1}
\begin{subequations}
\begin{equation}
\label{diffequ of u_l}
\left[-\frac{d^2}{dr^2}+\frac{l(l+1)}{r^2}-k^2+V(r)\right]u_l(r)=0,
\end{equation}
\begin{equation}
\label{diffequ of v_l}
\left[-\frac{d^2}{dr^2}+\frac{l(l+1)}{r^2}-k^2\right]v_l(r)=0,
\end{equation}
\end{subequations}
where $V(r)$ is a real, local, time-independent and short-range potential.

The asymptotic radial wave function $v_l$ takes the form
\begin{equation}
\label{v_l expression}
v_l(r)=e^{i\delta_l}i^l\frac{2l+1}{k}\left( \cos{\delta_l} \hat{j}_l(kr)+\sin{\delta_l} \hat{n}_l(kr) \right),
\end{equation}
where the phase shift $\delta_l$ is a function of $k$, and $\hat{j}$, $\hat{n}$ are the Riccati-Bessel functions of the first and second kinds~\cite{Math1}, respectively.

We now apply the Wronskian determinant to Eq.~\eqref{diffequ of u_l} by cross conjugating and multiplying the equations before subtraction, and derive
\begin{equation}
\label{wronski u_l 1}
-u_{2l}u^{\star\prime\prime}_{1l}+u^{\star}_{1l}u^{\prime\prime}_{2l}=(k_1^2-k_2^2)u^{\star}_{1l}u_{2l},
\end{equation}
where the subscripts 1 and 2 have been introduced to distinguish the two momenta $k_1$ and $k_2$, and primes denote derivatives with respect to $r$.

Integrating Eq.~\eqref{wronski u_l 1} from $r=\varepsilon$ to $\infty$ and simplifying the resulting expression by integration by parts gives:
\begin{equation}
\label{wronski u_l 2}
\left.u_{1l}^\star u_{2l}^\prime\right|_\varepsilon^\infty-\left.u_{2l}u_{1l}^{\star\prime}\right|_\varepsilon^\infty=(k_1^2-k_2^2)\int^\infty_\varepsilon u_{1l}^\star u_{2l}dr.
\end{equation}

Following an analogous procedure for Eq.~\eqref{diffequ of v_l}, we obtain a corresponding equation for $v_l$:
\begin{equation}
\label{wronski v_l 2}
\left.v_{1l}^\star v_{2l}^\prime\right|_\varepsilon^\infty-\left.v_{2l}^{} v_{1l}^{\star\prime}\right|_\varepsilon^\infty=(k_1^2-k_2^2)\int^\infty_\varepsilon v_{1l}^\star v_{2l}^{} dr.
\end{equation}

Subtracting Eq.~\eqref{wronski v_l 2} from Eq.~\eqref{wronski u_l 2} yields
\begin{equation}
\label{wronski minus}
\int_\varepsilon^\infty u_1^\star u_2-v_1^\star v_2 dr = \frac{\left.v_1^\star v_2^\prime - v_2 v_1^{\star\prime} + u_1^\star u_2^\prime - u_2 u_1^{\star\prime} \right|_{\varepsilon}}{k_1^2-k_2^2},
\end{equation}
where the subscript $l$ has been suppressed for brevity, and the boundary contributions at infinity from $u$ and $v$ cancel.

To evaluate the RHS of Eq.~\eqref{u2-v2}, we take the limit $k_1\to k_2$ in Eq.~\eqref{wronski minus}, which yields
\begin{equation}
\label{core formula nonCoulomb}
\int_\varepsilon^\infty \left| u \right|^2 - \left| v \right|^2 dr = \lim_{k_1\to k_2} \frac{\left.v_1^\star v_2^\prime\right|_\varepsilon - \left.v_1^{\star\prime} v_2\right|_\varepsilon}{k_1^2-k_2^2},
\end{equation}
where the terms involving $u$ vanish in the limit $\varepsilon\to 0$ as a consequence of the boundedness of the exact wave function.

Substituting Eq.~\eqref{v_l expression} into Eq.~\eqref{core formula nonCoulomb} gives the numerator as
\begin{subequations}
\begin{equation}
\label{v1*v2'}
\left.v_1^\star v_2^\prime\right|_\varepsilon = e^{i[\delta(k_2)-\delta(k_1)]}\frac{(2l+1)^2}{k_1}f_0(k_1)f_1(k_2),
\end{equation}
\begin{equation}
\label{v2v1*'}
\left.v_1^{\star\prime} v_2\right|_\varepsilon = e^{i[\delta(k_2)-\delta(k_1)]}\frac{(2l+1)^2}{k_2}f_0(k_2)f_1(k_1),
\end{equation}
\end{subequations}
where we have introduced auxiliary functions $f_n$ and the corresponding $g_n$ for later use, defined as
\begin{subequations}
\begin{equation}
\label{f_n}
f_n(k) = \cos{\delta}~\hat{j}^{(n)}(k\varepsilon) + \sin{\delta}~\hat{n}^{(n)}(k\varepsilon),
\end{equation}
\begin{equation}
\label{g_n}
g_n(k) = \cos{\delta}~\hat{n}^{(n)}(k\varepsilon) - \sin{\delta}~\hat{j}^{(n)}(k\varepsilon),
\end{equation}
\end{subequations}
with the superscript $(n)$ denoting the n-th derivative with respect to $kr$. These derivatives can be derived via recurrence relations of Riccati-Bessel funtions~\cite{Math1}:
\begin{equation}
\label{recurrence}
\chi_l^\prime=\chi_{l-1}-\frac{l}{z}\chi_l,
\end{equation}
where $\chi$ denotes either $\hat{j}$ or $\hat{n}$.

The numerator $N$ in the limiting expression on the RHS of Eq.~\eqref{core formula nonCoulomb}, after the use of L'Hospital's rule becomes
\begin{widetext}
\begin{equation}
\label{Numerator diff}
\frac{dN}{dk_1}=e^{i[\delta(k_2)-\delta(k_1)]}(2l+1)^2\left\{\frac{d}{dk_1}\left[ \frac{f_0(k_1)f_1(k_2)}{k_1} -\frac{f_0(k_2)f_1(k_1)}{k_2} \right]-i\frac{d\delta(k_1)}{dk_1}\left[ \frac{f_0(k_1)f_1(k_2)}{k_1} -\frac{f_0(k_2)f_1(k_1)}{k_2} \right]\right\}.
\end{equation}
\end{widetext}
The second term in the curly-bracketed expression of Eq.~\eqref{Numerator diff} tends to zero as $k_1\to k_2$. The functions $f_n$ appearing under derivatives can be substituted using the relation
\begin{equation}
\label{diff f_n}
\frac{df_n(k)}{dk} = \frac{d\delta(k)}{dk}g_n(k)+rf_{n+1}(k),
\end{equation}
so that Eq.~\eqref{Numerator diff} can be further evaluated and, after taking the limit $k_1\to k_2$ with $k_2=k$, reduces to
\begin{eqnarray}
\label{Numerator diff 2}
\frac{dN}{dk_1}\to (2l+1)^2&&\left[ \frac{1}{k}\frac{d\delta}{dk}+\frac{\varepsilon}{k}\left(f_1^2-f_0^2\left(\frac{l(l+1)}{(kr)^2}-1\right)\right)\right.\nonumber \\
&&\left.-\frac{1}{k^2}f_0f_1 \right].
\end{eqnarray}

Now we have completed the evaluation of the RHS of Eq.~\eqref{core formula nonCoulomb}, as well as the corresponding term in Eq.~\eqref{newCF}. While theoretically feasible, the present formulation is numerically problematic because the first subtraction in Eq.~\eqref{newCF} involves two large quantities that nearly cancel when $\varepsilon\to 0$, leading to severe rounding errors. 

To circumvent this issue, we extract the divergent parts analytically and perform the subtraction before any numerical evaluation. As $\varepsilon\to 0$, the Riccati–Bessel functions $\hat{j}_l(k\varepsilon)$ and $\hat{n}_l(k\varepsilon)$ behave as
\begin{equation}
\label{j n at 0}
\hat{j}_l(k\varepsilon) \sim \frac{2^{-l-1}\sqrt{\pi}(k\varepsilon)^{l+1}}{\Gamma (l+\frac{3}{2})}
~~~;~~~
\hat{n}_l(k\varepsilon) \sim \frac{2^{l}\Gamma (l+\frac{1}{2})}{\sqrt{\pi}(k\varepsilon)^l}
\end{equation}

We isolate the divergent part in Eq.~\eqref{core formula nonCoulomb} and define another auxiliary function $f$ such that its integration satisfies:
\begin{eqnarray}
\label{func f}
\int_\varepsilon^\infty f(r)dr  &&= \frac{(2l+1)^2}{2k^3}\sin^2\delta \left[ k\varepsilon\left( \hat{n}_l^2+\hat{n}_{l-1}^2 \right)\right.\nonumber\\
&&\left.-(2l+1)\hat{n}_l\hat{n}_{l-1} \right]\theta(\lambda-\varepsilon),
\end{eqnarray}
where a step function $\theta$ has been inserted, with the free parameter $\lambda>\varepsilon$.

The function $f$ can be obtained from Eq.~\eqref{func f} by differentiating with respect to $\varepsilon$:
\begin{eqnarray}
\label{func f 2}
f(r)&&=\frac{(2l+1)^2}{2k^3}\sin^2\delta\left\{-2k\hat{n}_l^2\theta(\lambda-r)\right.\nonumber\\
&&+\left.\left[ kr\left( \hat{n}_l^2+\hat{n}_{l-1}^2 \right)-\left( 2l+1 \right)\hat{n}_l\hat{n}_{l-1} \right]\delta(\lambda-r)
\right\}.~~~~
\end{eqnarray}
The first term in the curly brackets exactly cancels the divergent part of $|v|^2$ in Eq.~\eqref{core formula nonCoulomb}, leaving only the finite remainder. Consequently, one may safely take $\varepsilon=0$ without encountering any singularity.

The remaining finite terms, denoted collectively by $C$, are
\begin{eqnarray}
\label{Convergent terms}
C && = \frac{(2l+1)^2}{2k^2}\left\{ -\frac{(2l+1)\sin(2\delta)}{2k}+\frac{d\delta}{dk} \right. \nonumber\\
           && + \left. \frac{\sin^2\delta}{k}\left[ k\lambda\left( \hat{n}_l^2+\hat{n}_{l-1}^2 \right)-(2l+1)\hat{n}_l\hat{n}_{l-1} \right] \right\},
\end{eqnarray}
where $\hat{n}_i$ is evaluated at $k\lambda$.

With the preparation above, we now return to Eq.~\eqref{newCF} and formulate our approach to the correlation function. The functions $\Phi$ and $v_l$ in Eqs.~\eqref{newCF} and \eqref{Phi_partialwave} are replaced by modified versions $\Tilde{\Phi}$ and $\Tilde{v}_l$, defined as
\begin{equation}
\Tilde{\Phi}=\sum_{l=0}^\infty \Tilde{v}_l(r)P_l(\cos{\theta}),
\end{equation}
\begin{equation}
\Tilde{v}_l(r)=e^{i\delta_l}i^l\frac{2l+1}{k}\left( \cos{\delta_l} \hat{j}_l(kr)+\sin{\delta_l} \Tilde{n}_l(kr) \right),
\end{equation}
where $\Tilde{n}_l(kr)\coloneqq\hat{n}_l(kr)\theta(r-\lambda)$.

Then the new partial wave correction term can be written as
\begin{eqnarray}
\label{new correction term}
&&\int_0^\infty \left| u \right|^2-\left| \Tilde{v} \right|^2 dr \nonumber \\
&& = \int_0^\infty \left| u \right|^2 - \left( \left|v\right|^2-\frac{(2l+1)^2}{k^2}\sin^2\delta\hat{n}_l^2\theta(\lambda-r) \right)   dr  \nonumber \\
&& +\frac{(2l+1)^2}{k^3}\sin(2\delta)\int_0^{k\lambda}\hat{j}_l\hat{n}_l d(kr).
\end{eqnarray}
The first term on the RHS of Eq.~\eqref{new correction term} is given by Eq.~\eqref{Convergent terms}, while the second admits a closed-form analytic integration:
\begin{eqnarray}
\label{int jn dr}
&&\int_0^{k\lambda}\hat{j}_l\hat{n}_ldr\nonumber \\
&&= \frac{2l+1}{4}\left( 1-{}_1F_2\left( -\frac{1}{2};-l-\frac{1}{2},l+\frac{1}{2};-k^2\lambda^2 \right) \right),~~~
\end{eqnarray}
where ${}_1F_2$ denotes the Hypergeometric function~\cite{Math1}.

According to the effective range expansion, the undetermined quantities $\delta_l$ and $d\delta_l/dk$ can be parametrized by~\cite{EffRange_nc}
\begin{equation}
\label{ERE_nC}
k^{2l+1}\cot{\delta_l(k)}=\sum_{i=0}^\infty C_{i,l} k^{2i},
\end{equation}
where the scattering length is $a_l=-1/C_{0,l}$, the effective range is $d_l=2C_{1,l}$ and the form factors are related to the higher-order coefficients. In most circumstances, only the first few $C_{i,l}$ are needed to describe the correlation function accurately.\vspace{0pt}

We have now completed the construction of $C(\textbf{k},\textbf{K})$ in terms of the coefficients $C_{i,l}$, which are determined solely by the potential $V(r)$. By fitting these coefficients, we can extract essential information about the interaction. The free parameter $\lambda$ can vary over a broad range, bounded from above by the approximation discussed after Eq.~\eqref{newCF} and from below by floating-point precision. It can also take on distinct values for different values of $k$ and $l$.
\subsection{The Coulomb case}
The procedure for the Coulomb case is analogous but more involved. The radial Schrödinger equations read~\cite{QM1,QM2}
\begin{subequations}
\begin{equation}
\label{diffequ of u_l C}
\left[-\frac{d^2}{dr^2}+\frac{l(l+1)}{r^2}-k^2+V(r)+\frac{2k\eta}{r}\right]u_l(r)=0,
\end{equation}
\begin{equation}
\label{diffequ of v_l C}
\left[-\frac{d^2}{dr^2}+\frac{l(l+1)}{r^2}-k^2+\frac{2k\eta}{r}\right]v_l(r)=0,
\end{equation}
\end{subequations}
where the Sommerfeld parameter is $\eta=1/(a_ck)$, with $a_c$ the Coulomb scattering length. The Coulomb analogue of $v_l$ differs from Eq.~\eqref{v_l expression} and takes the form
\begin{equation}
\label{v_l expression C}
v_l=e^{i\left( \delta_l+\sigma_l \right)}i^l\frac{2l+1}{k}\left[ \cos\delta_lF_{l,\eta}(kr)+\sin\delta_lG_{l,\eta}(kr) \right],
\end{equation}
where $\sigma_l$ is the Coulomb phase shift, equal to $\arg \Gamma(l+1+i\eta)$, and $F$ and $G$ are the regular and irregular Coulomb wave functions~\cite{Math1,Math2}, respectively.

Owing to the momentum dependence of $\eta$, the derivatives of the auxiliary functions $f_n$ acquire an additional term:
\begin{equation}
\label{aux f_n C}
\frac{df_n(k)}{dk}=\frac{d\delta(k)}{dk}g_n(k)+rf_{n+1}(k)-\frac{\eta}{k}\frac{\partial f_n(k)}{\partial\eta}.
\end{equation}

The derivatives that appear can also be evaluated using the recurrence relations of the Coulomb wave functions~\cite{Math1}:
\begin{equation}
\label{recurrence C}
\chi^\prime=\sqrt{1+\frac{\eta^2}{l^2}}\chi_{l-1}-\left( \frac{l}{\rho}+\frac{\eta}{l} \right)\chi_l,
\end{equation}
where $\chi$ denotes either $F$ or $G$. At first sight, Eq.~\eqref{recurrence C} is ill-defined at $l=0$, which would require separate treatment of this case. In fact, however, the final result remains well-defined and holds for all $l$. We therefore omit the separate check for $l=0$.

Then we follow the same procedure as in the non-Coulomb case, up to the point where L'Hôpital's rule is applied to the Coulomb version of Eq.~\eqref{core formula nonCoulomb}. The momentum dependency of $\sigma_l$ does not introduce any additional term here.
The nontrivial part in the derivative of the numerator $N$ is expanded as
\begin{widetext}
\begin{equation}
\label{d/dk (f0f1/k-f0f1/k)}
\frac{d}{dk_1}\left[ \frac{f_0(k_1)f_1(k_2)}{k_1}-\frac{f_0(k_2)f_1(k_1)}{k_2}\right]=\frac{1}{k}\frac{d\delta}{dk}+\frac{r}{k}\left( f_1^2-f_0f_2 \right)-\frac{f_0f_1}{k^2}-\frac{\eta}{k}\left(  \frac{\partial f_0}{\partial \eta}f_1-\frac{\partial f_1}{\partial \eta}f_0 \right).
\end{equation}
\end{widetext}

After isolating the divergent terms, the auxiliary function $f$ takes the form
\begin{eqnarray}
\label{func f C}
f(r)&&=\frac{(2l+1)^2}{2k^3}\sin^2\delta\left\{-2kG_l^2\theta(\lambda-r)\right.\nonumber\\
&&+ \left[krG_l^{\prime 2}+(kr-2\eta)G_l^2-G_lG_l^\prime-\frac{l(l+1)}{kr}G_l^2\right.\nonumber\\
&&\left.\left. +\eta\left( G_l^\prime\frac{\partial G_l}{\partial \eta}-G_l\frac{\partial G_l^\prime}{\partial \eta} \right) \right]\delta(\lambda-r)  \right\}.~~~~
\end{eqnarray}
The first term in the curly brackets again cancels the divergent part of $|v|^2$. The remaining finite terms, denoted collectively by $C$, are
\begin{eqnarray}
\label{Convergent terms C}
C &&= \frac{(2l+1)^2}{2k^2}\left\{  \frac{d\delta}{dk}-\frac{\sin(2\delta)}{2k}\sqrt{1+\frac{\eta^2}{l^2}}\frac{C_{l-1}}{C_l}\left(1\right. \right. \nonumber\\
           &&\left.-\frac{\eta}{2l+1}\left( \frac{C_l^\prime}{C_l}+\frac{C_{l-1}^\prime}{C_{l-1}}+\frac{\eta}{l^2+\eta^2}\right)\right)\nonumber\\
           &&+ \frac{\sin^2\delta}{k}\left[krG_l^{\prime 2}+(kr-2\eta)G_l^2-G_lG_l^\prime-\frac{l(l+1)}{kr}G_l^2\right.\nonumber\\
           &&\left. +\eta\left( G_l^\prime\frac{\partial G_l}{\partial \eta}-G_l\frac{\partial G_l^\prime}{\partial \eta} \right) \right]\delta(\lambda-r)\nonumber\\
           &&\left.+2\frac{\sin(2\delta)}{k}\int_0^{k\lambda}F_lG_ld(kr)\right\},
\end{eqnarray}

where $C_n$ is the normalization constant for the Coulomb wave functions, defined by~\cite{Math1}
\begin{equation}
\label{C_i expression}
C_n=\frac{2^ne^{-\pi\eta/2}\left| \Gamma(n+1+i\eta) \right|}{(2n+1)!}.
\end{equation}
Thus, the second term in the curly brackets, denoted by $T$, can be further simplified to
\begin{eqnarray}
\label{C_i simplify}
T&&=-\frac{\sin(2\delta)}{2k^2}(2l+1)\left( 1-\frac{\eta}{2l+1}\left( -\pi\right.\right. \nonumber \\
&&\left.\left.+\frac{2\eta}{l^2+\eta^2}+i\psi^{(0)}(l+i\eta)-i\psi^{(0)}(l-i\eta) \right) \right).
\end{eqnarray}

Again using the effective range expansion, the phase shift and its derivative in the Coulomb case are given by~\cite{ERE_C}
\begin{equation}
\label{ERE_C}
C_l^2 k^{2l+1} \cot\delta_l+2k\eta h_l(k) = \sum_{i=0}^\infty C_{i,l} k^{2i},
\end{equation}
where $h_l(k)$ is defined as
\begin{equation}
h_l(k)=k^{2l}\frac{C_l^2}{C_0^2}\left\{ \re{\psi(i\eta)}-\ln{|\eta|} \right\}.
\end{equation}

\section{Results and discussions}
In this section, we verify the method from several perspectives and explore the limits of its applicability. The physical systems we consider are spin-0 neutron–neutron and proton–proton pairs, described by the Reid soft-core potential including $V({}^1S_0)$ and $V({}^1D_2)$ channels~\cite{ReidPotential}. In order to avoid the non-central interaction, spin-1 pairs are excluded from consideration. In the following calculations, the actual energy spectra of the emitted nucleons are neglected, so that the total momentum $K$ of each pair is set to zero accordingly.

The phase shift $\delta_l$ and the exact wave function are computed using the variable phase method and its extension~\cite{VPEextension}. Reference values of the coefficients in the effective range expansion are obtained by fitting the $\delta_l(k)$ versus $k$ relation.

\begin{figure}
    \centering
    \includegraphics[width=0.4\textwidth]{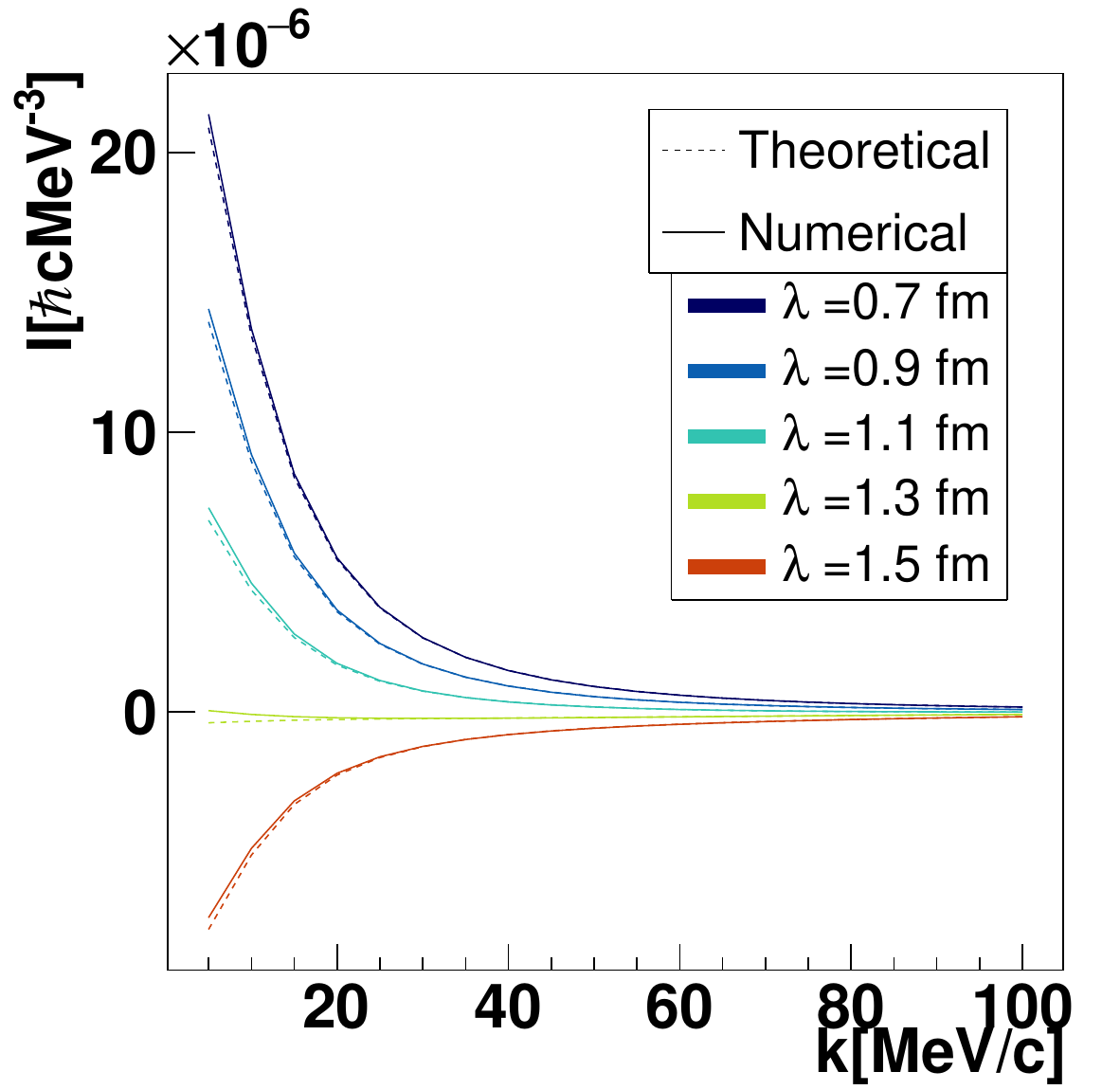}
    \caption{Verification of Eq.~\eqref{core formula nonCoulomb} for $l=0$ in the non-Coulomb case, comparing the numerical LHS with the analytical RHS. The quantity $I$ denotes the value of the integral.}
    \label{fig:u2-v2 1}
\end{figure}
\begin{figure}
    \centering
    \includegraphics[width=0.4\textwidth]{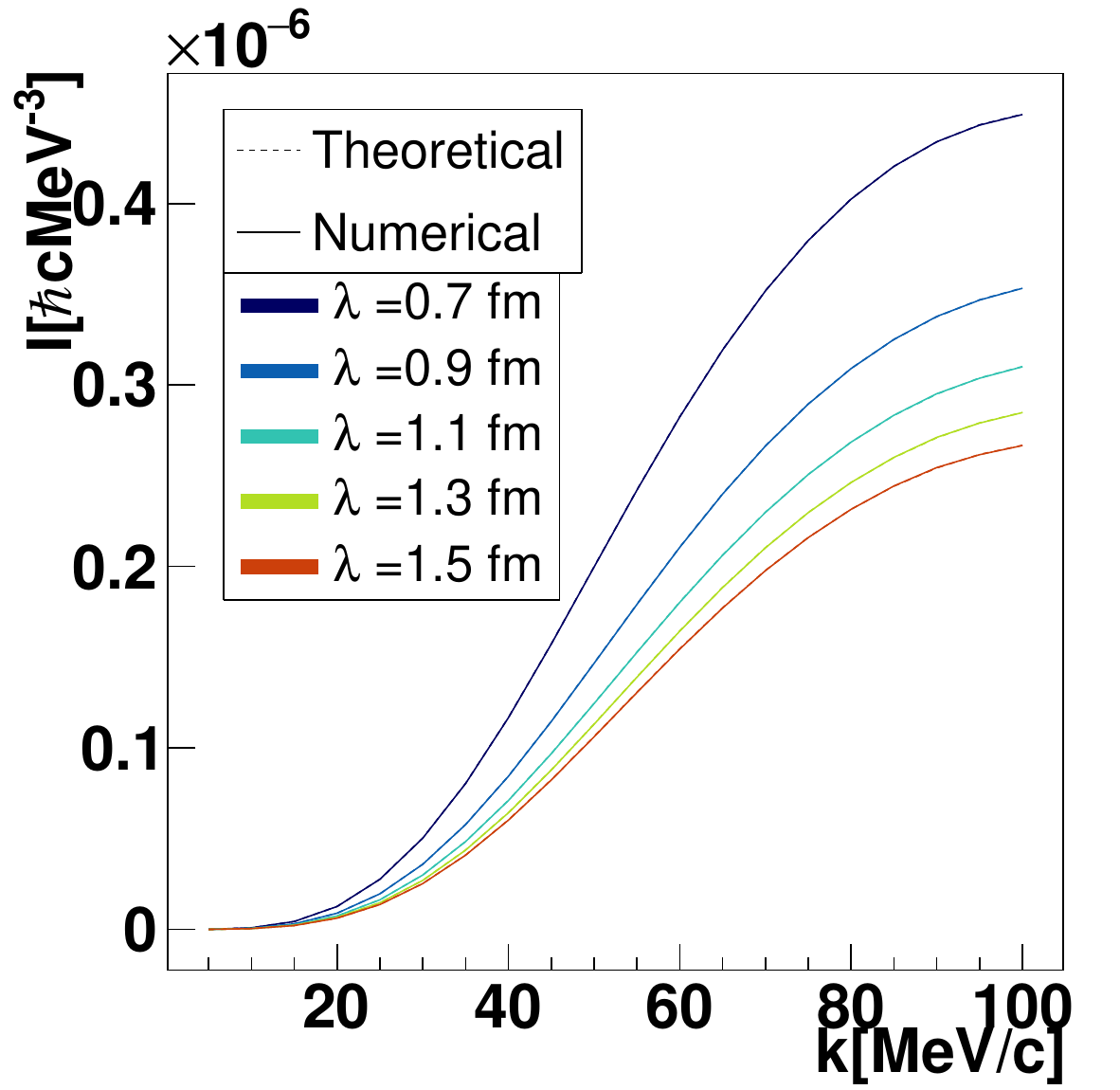}
    \caption{Verification of Eq.~\eqref{core formula nonCoulomb} for $l=2$ in the non-Coulomb case, comparing the numerical LHS with the analytical RHS.}
    \label{fig:u2-v2 2}
\end{figure}
\begin{figure}
    \centering
    \includegraphics[width=0.4\textwidth]{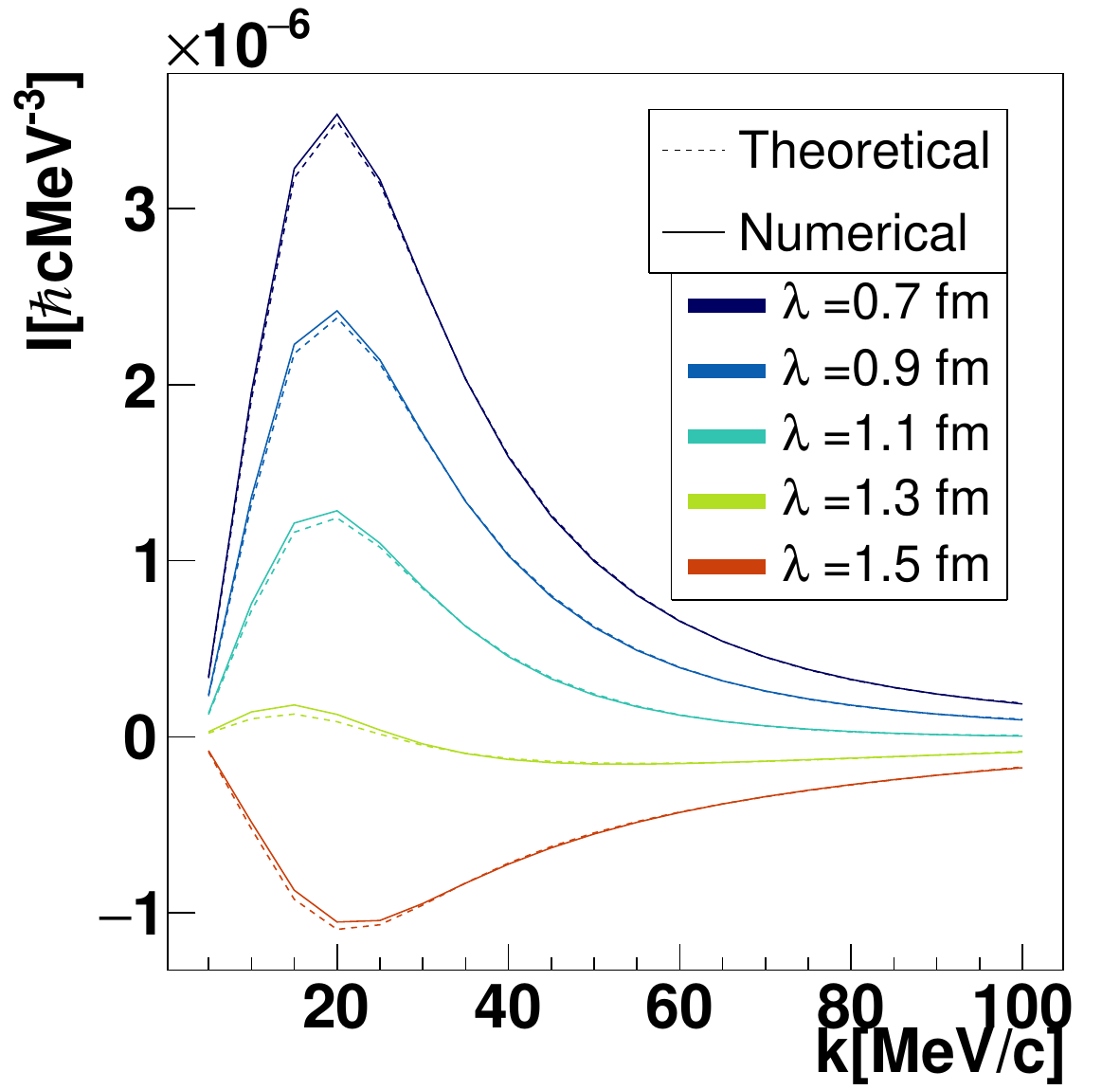}
    \caption{Verification of Eq.~\eqref{core formula nonCoulomb} for $l=0$ in the Coulomb case, comparing the numerical LHS with the analytical RHS.}
    \label{fig:u2-v2 3}
\end{figure}

\begin{figure}
    \centering
    \includegraphics[width=0.4\textwidth]{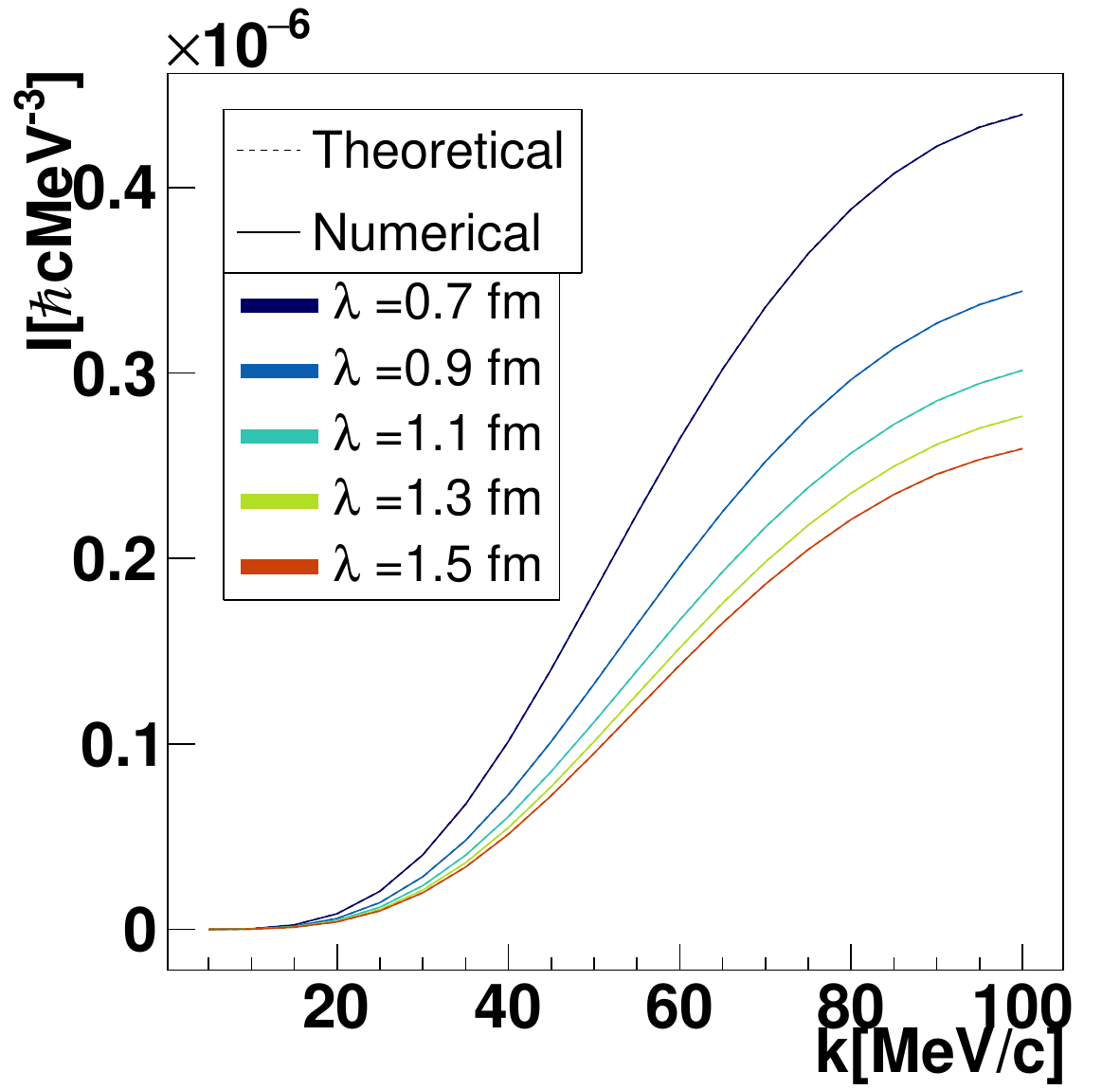}
    \caption{Verification of Eq.~\eqref{core formula nonCoulomb} for $l=2$ in the Coulomb case, comparing the numerical LHS with the analytical RHS.}
    \label{fig:u2-v2 4}
\end{figure}

\subsection{Verification of the radial correction terms}

Prior to generating correlation functions with our new method, we validate the core formula in Eq.~\eqref{core formula nonCoulomb} and its Coulomb counterpart. The LHS of Eq.~\eqref{core formula nonCoulomb}, which involves the exact wave function $u$, is evaluated numerically, while the RHS is computed analytically from the results derived above. Both partial waves $l=0$ and $2$ are examined for several values of the free parameter $\lambda$.

\begin{figure}
    \centering
    \includegraphics[width=0.4\textwidth]{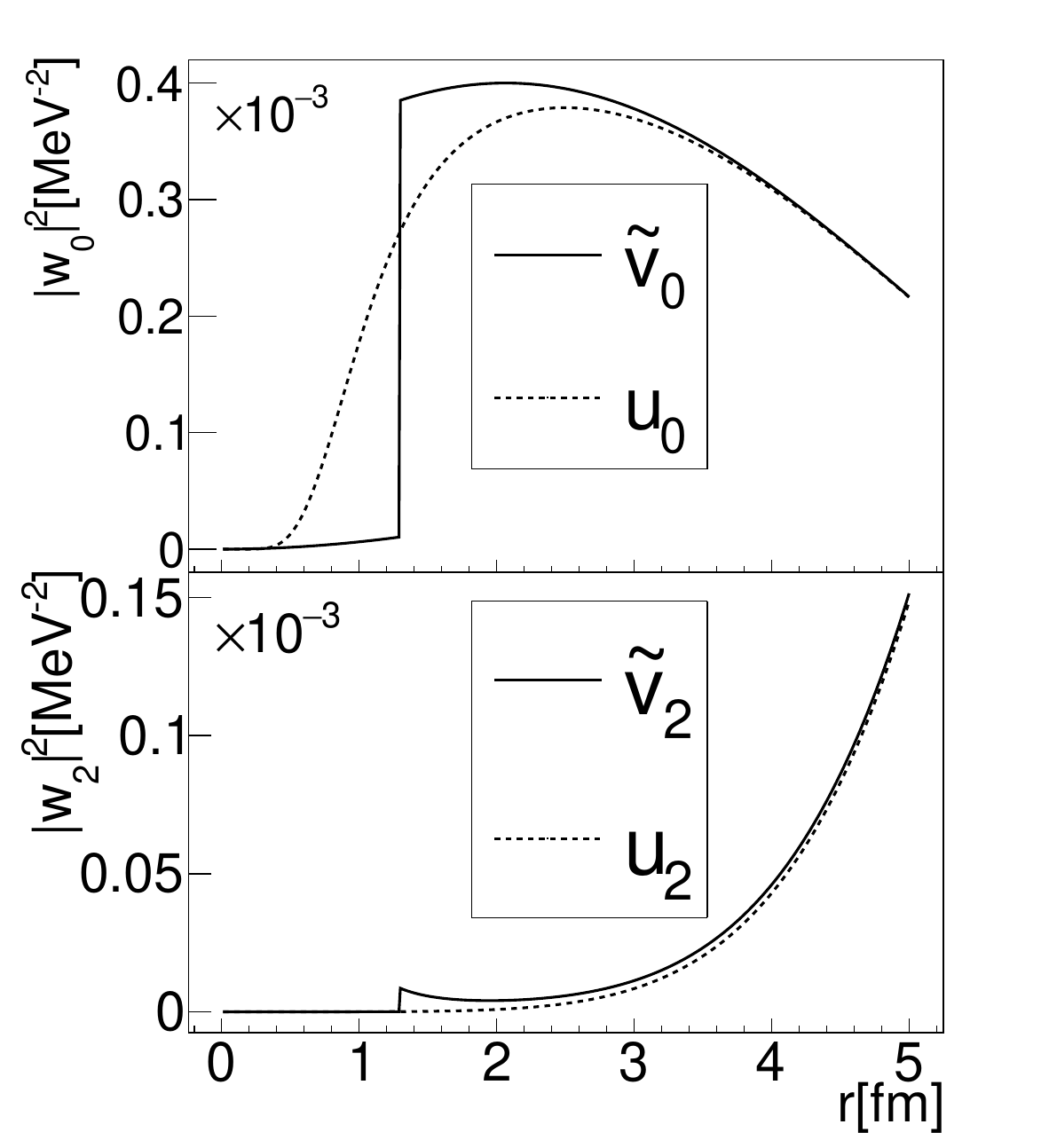}
    \caption{Square of the modulus of the exact wave function $u_l$ and of the modified wave function $\Tilde{v}_l$ at $\lambda$= 1.3\,fm and $k=50$\,MeV}
    \label{fig:u2 v2 - r}
\end{figure}
From Figs.\ref{fig:u2-v2 1} to \ref{fig:u2-v2 4}, the overall agreement between the theoretical and numerical curves is excellent for different values of $\lambda$ and $k$, convincingly validating our analytical calculation.

Beyond this agreement, the $\lambda$ dependency of the curves also reveals the behavior of $u_l$ near the origin. In Figs.\ref{fig:u2-v2 1} and \ref{fig:u2-v2 3}, when $\lambda\approx1.3$\,fm, $I$ is almost zero for different momenta, which indicates that $|u_0|^2$ falls rapidly to zero at the origin, as shown in Fig.\ref{fig:u2 v2 - r}. For $l=2$, however, values of $\lambda$ over a wide range all lead to a significant $I$, which reflects the smooth behavior of $|u_2|^2$ near the origin seen in Fig.~\ref{fig:u2 v2 - r}.

A comparison between the Coulomb and non-Coulomb cases reflects the screening of short-range forces, such as the nuclear force, by the Coulomb force. At low momentum $k$, the Coulombian correction term is strongly suppressed and vanishes, as shown in Fig.\ref{fig:u2-v2 3}. At higher $k$, this screening effect diminishes, so that Figs.\ref{fig:u2-v2 2} and \ref{fig:u2-v2 4} become quite similar.

\subsection{Generation of correlation functions}

To estimate the systematic uncertainty introduced by the approximation in Eq.~\eqref{newCF}, we need to generate a series of neutron-neutron and proton-proton correlation functions for different Gaussian source sizes $\sigma$. Before applying our method, we need the coefficients in Eqs.~\eqref{ERE_nC} and \eqref{ERE_C}, for which the infinite summation is truncated to the first three terms. The free parameter $\lambda$ is fixed at 1\,fm throughout.

\begin{figure}
    \centering
    \includegraphics[width=0.4\textwidth]{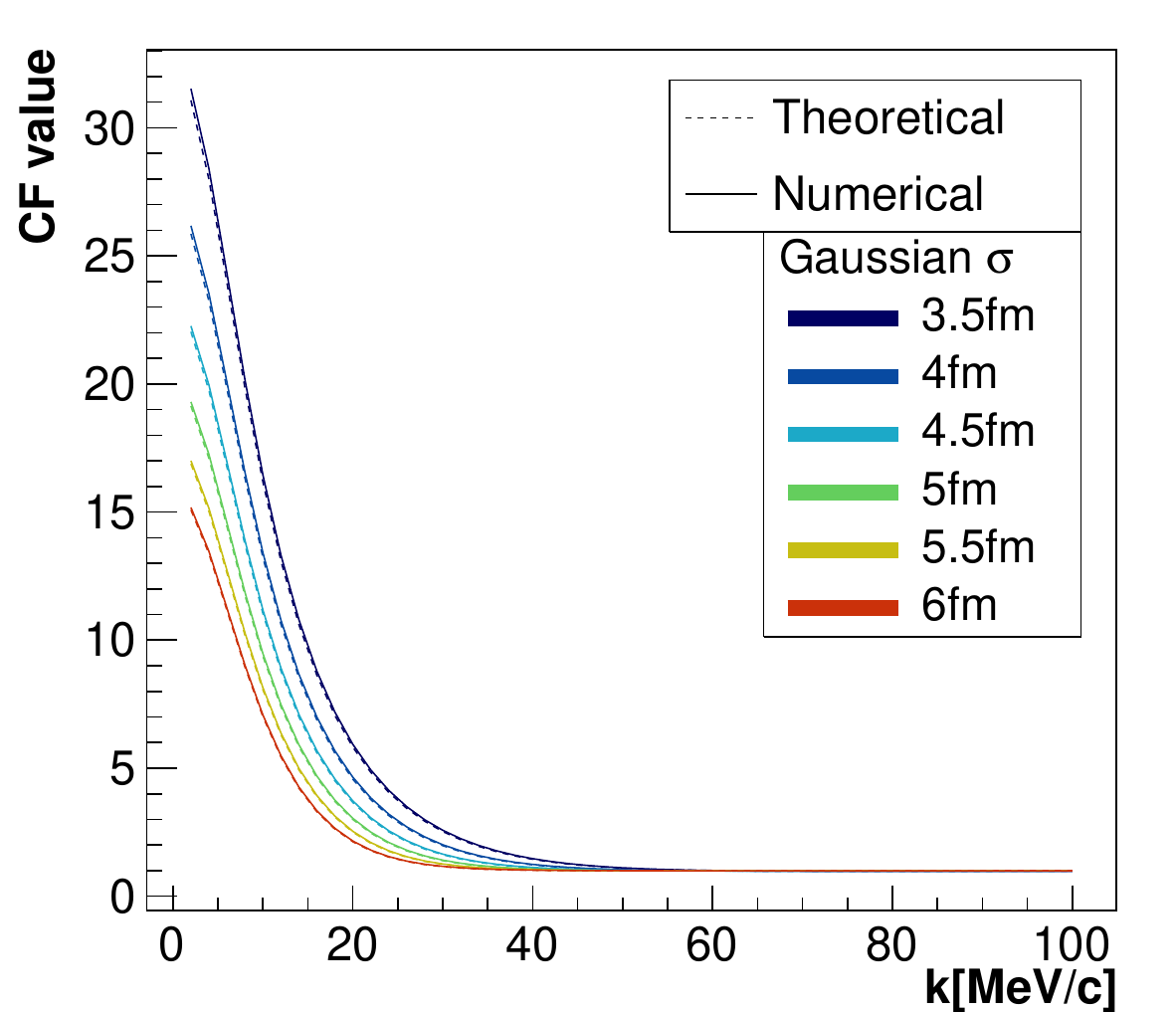}
    \caption{Neutron-neutron correlation functions for the $S=0$ channel with different Gaussian source sizes $\sigma$. The numerical curves are computed from the exact wave function, while the theoretical curves are obtained with our method.}
    \label{fig:nn CF}
\end{figure}

\begin{figure}
    \centering
    \includegraphics[width=0.4\textwidth]{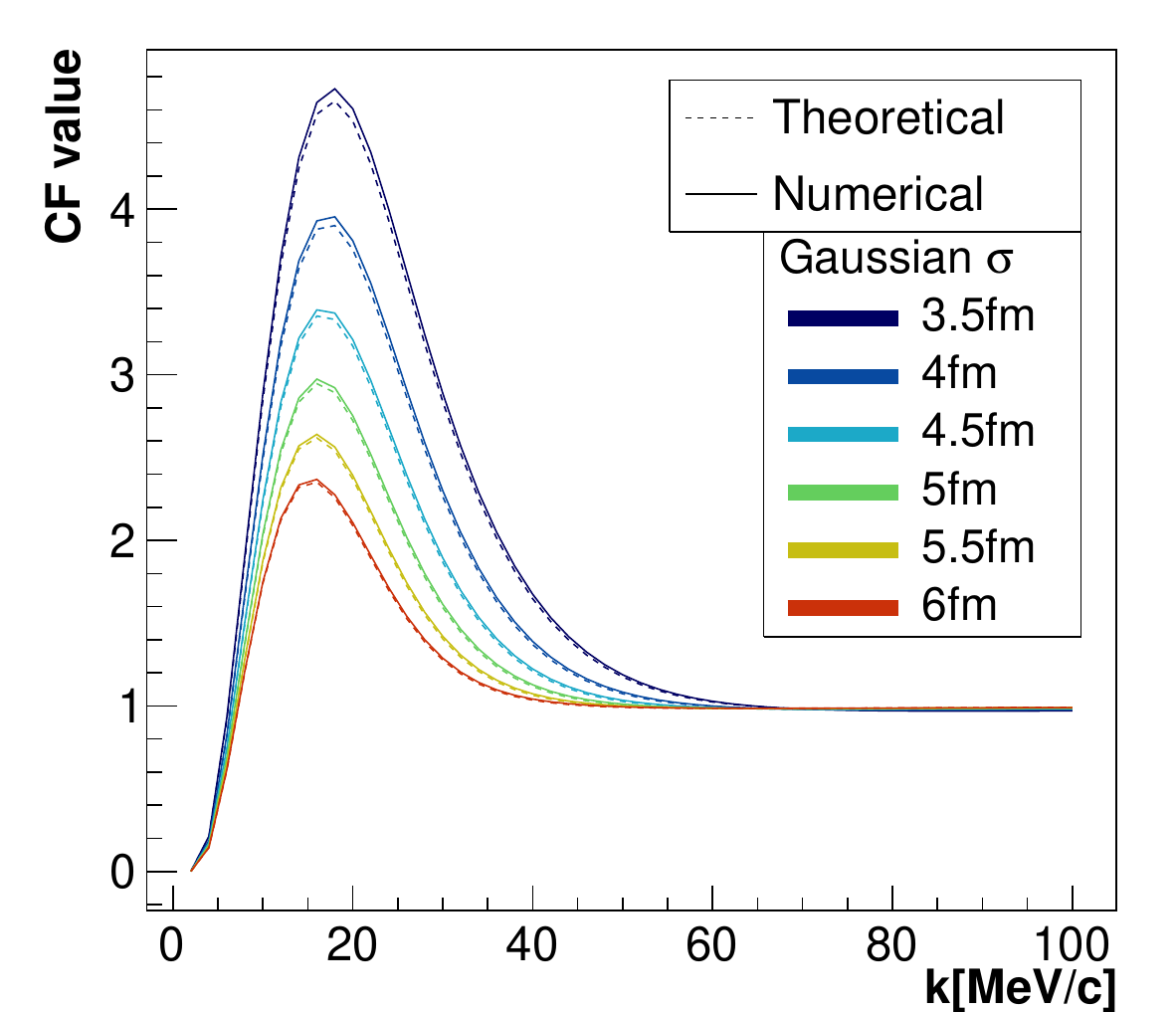}
    \caption{Proton-proton correlation functions for the $S=0$ channel with different $\sigma$.}
    \label{fig:pp CF}
\end{figure}

As shown in Figs.\ref{fig:nn CF} and \ref{fig:pp CF}, the difference between the theoretical and numerical curves clearly decreases as $\sigma$ increases. This behavior originates from the approximation in Eq.~\eqref{newCF}, which is better satisfied for a larger Gaussian source.

\begin{figure}
    \centering
    \includegraphics[width=0.4\textwidth]{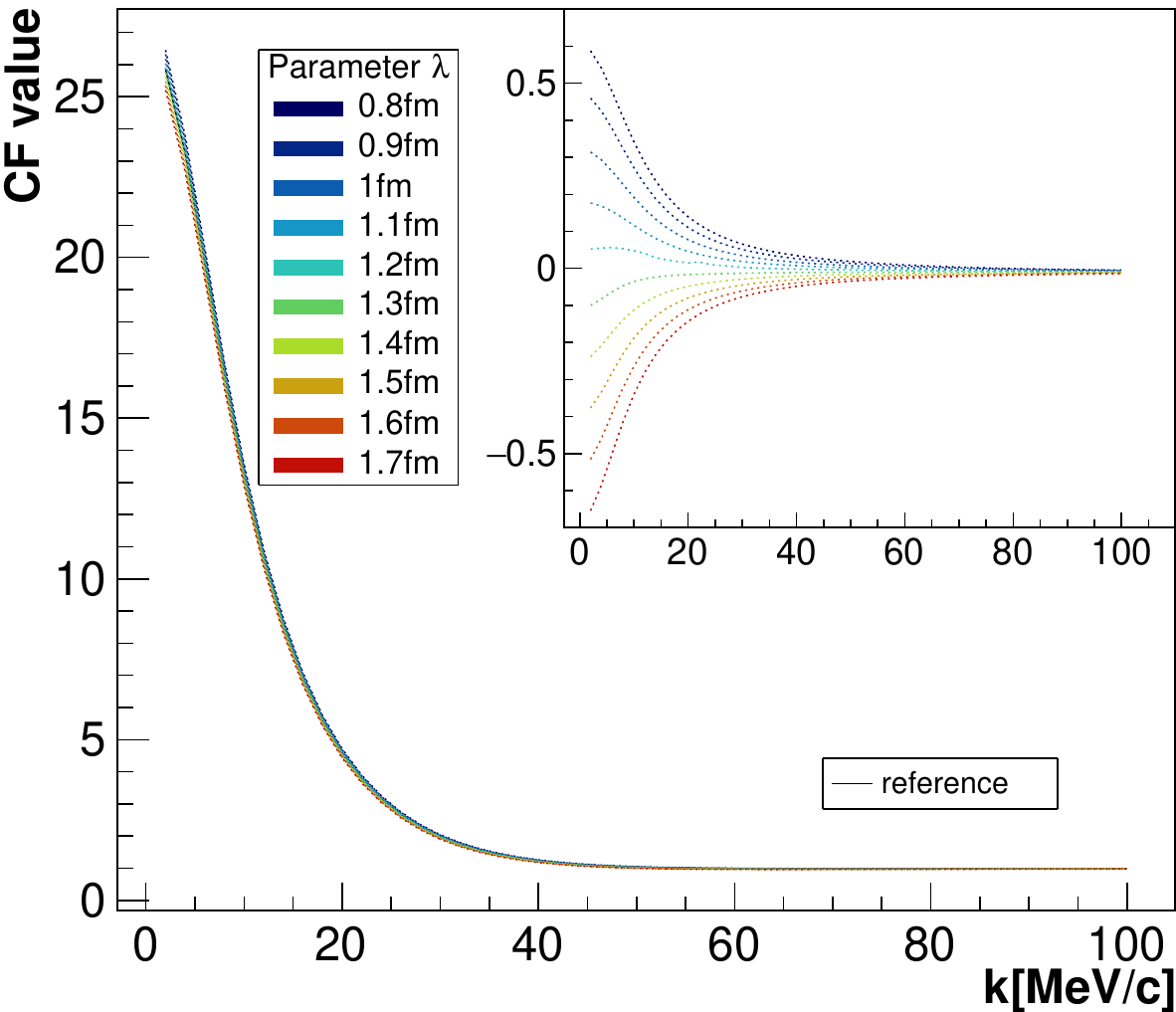}
    \caption{Neutron-neutron correlation functions for the $S=0$ channel with different values of $\lambda$. The reference curve is computed from the exact wave function; deviations from it are shown in the inset.}
    \label{fig:nnVF}
\end{figure}

\begin{figure}
    \centering
    \includegraphics[width=0.4\textwidth]{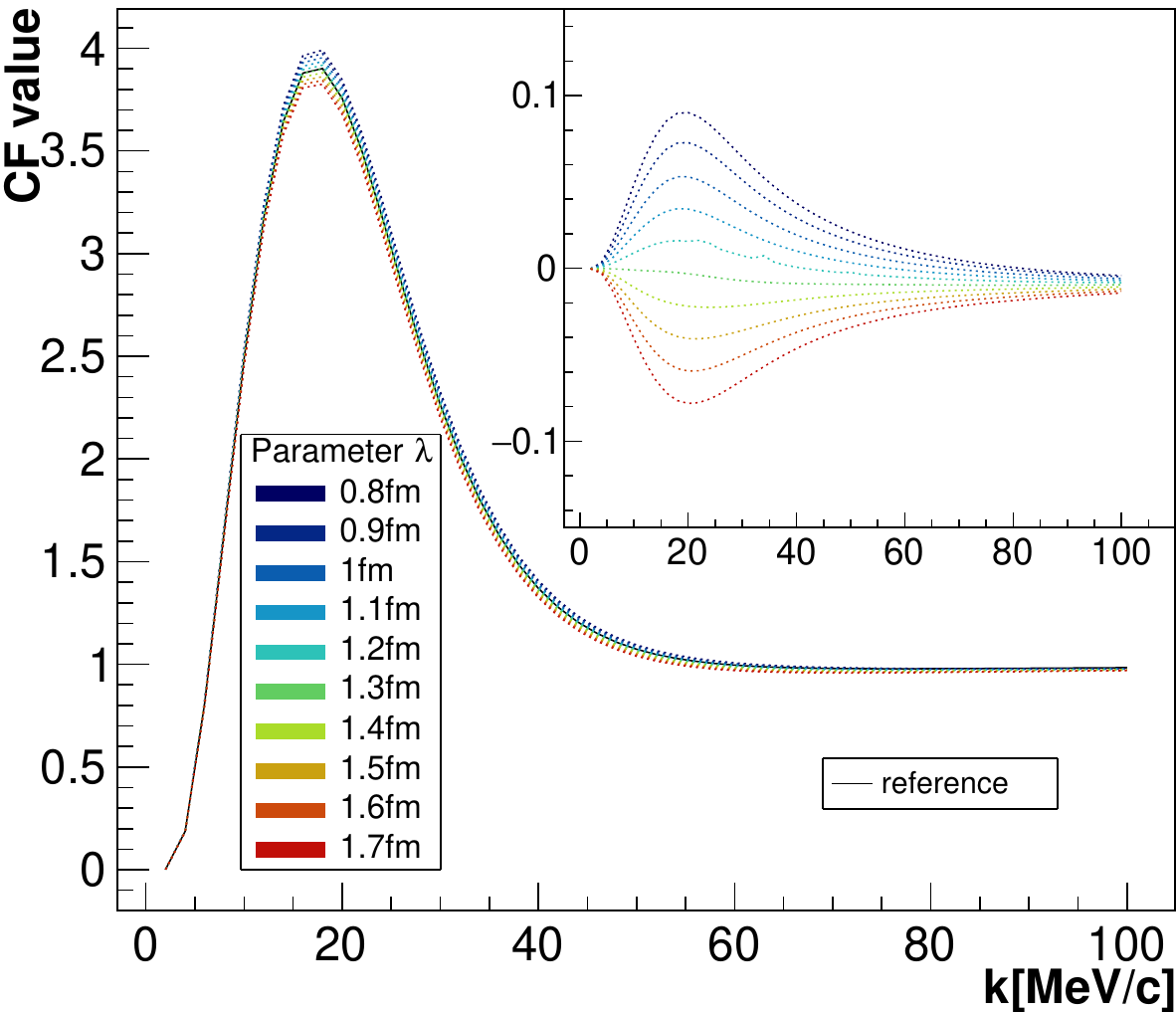}
    \caption{Proton-proton correlation functions for the $S=0$ channel with different values of $\lambda$.}
    \label{fig:ppVF}
\end{figure}

To further investigate the influence of $\lambda$ on the correlation functions, we fix $\sigma$ at 4\,fm and vary $\lambda$, obtaining Figs.\ref{fig:nnVF} and \ref{fig:ppVF}. 

When $\lambda\approx1.3$\,fm, the deviation from the reference curve reaches its minimum, consistent with the value of $\lambda$ at which $I\approx0$. Therefore, to minimize the model uncertainty, one may choose a specific $\lambda$ for each momentum bin so that the correction terms tend to vanish, thereby suppressing the influence of the approximation. The deviation is also positively correlated with the magnitude of the correlation function. In most experiments, the model uncertainty introduced here remains acceptable.
\subsection{Extraction of interaction parameters}

In this subsection, we apply our method to a simulated neutron-neutron correlation function in the spin-0 channel and extract the interaction parameters of s and d-waves. The simulated correlation function is generated from the exact wave function with a Gaussian source of $\sigma=4$\,fm. To better assess the ability of our method to extract higher partial wave interactions, we artificially enhance the d-wave interaction by a factor of 10. Otherwise, its contribution would be buried in noise.

To accelerate the fitting procedure, we preprocess the coefficients in Eq.~\eqref{ERE_nC}, since their magnitudes differ greatly and are not readily amenable to most fitting algorithms. We rescale the coefficients via $k\to k/200$ (natural units), and scale down the s-wave coefficients by a factor of $10^2$ and the d-wave coefficients by a factor of $10^{11}$. Consequently, the interaction parameters $p_{l,j}$ involved in the fit can be expressed as
\begin{subequations}
\begin{equation}
\label{s-wave fit}
k\cot{\delta_0}\approx10^2\left[ p_{0,0}+p_{0,1}\left(\frac{k}{200}\right)^2 \right],
\end{equation}
\begin{eqnarray}
\label{d-wave fit}
k^5\cot{\delta_2}&&\approx10^{11}\left[ p_{2,0}+p_{2,1}\left(\frac{k}{200}\right)^2\right.\nonumber\\
&&\left.+p_{2,2}\left(\frac{k}{200}\right)^4+p_{2,3}\left(\frac{k}{200}\right)^6 \right].
\end{eqnarray}
\end{subequations}
All the coefficients of interest now have magnitudes of order 1 or 0.1.

\begin{figure}
    \centering
    \includegraphics[width=0.4\textwidth]{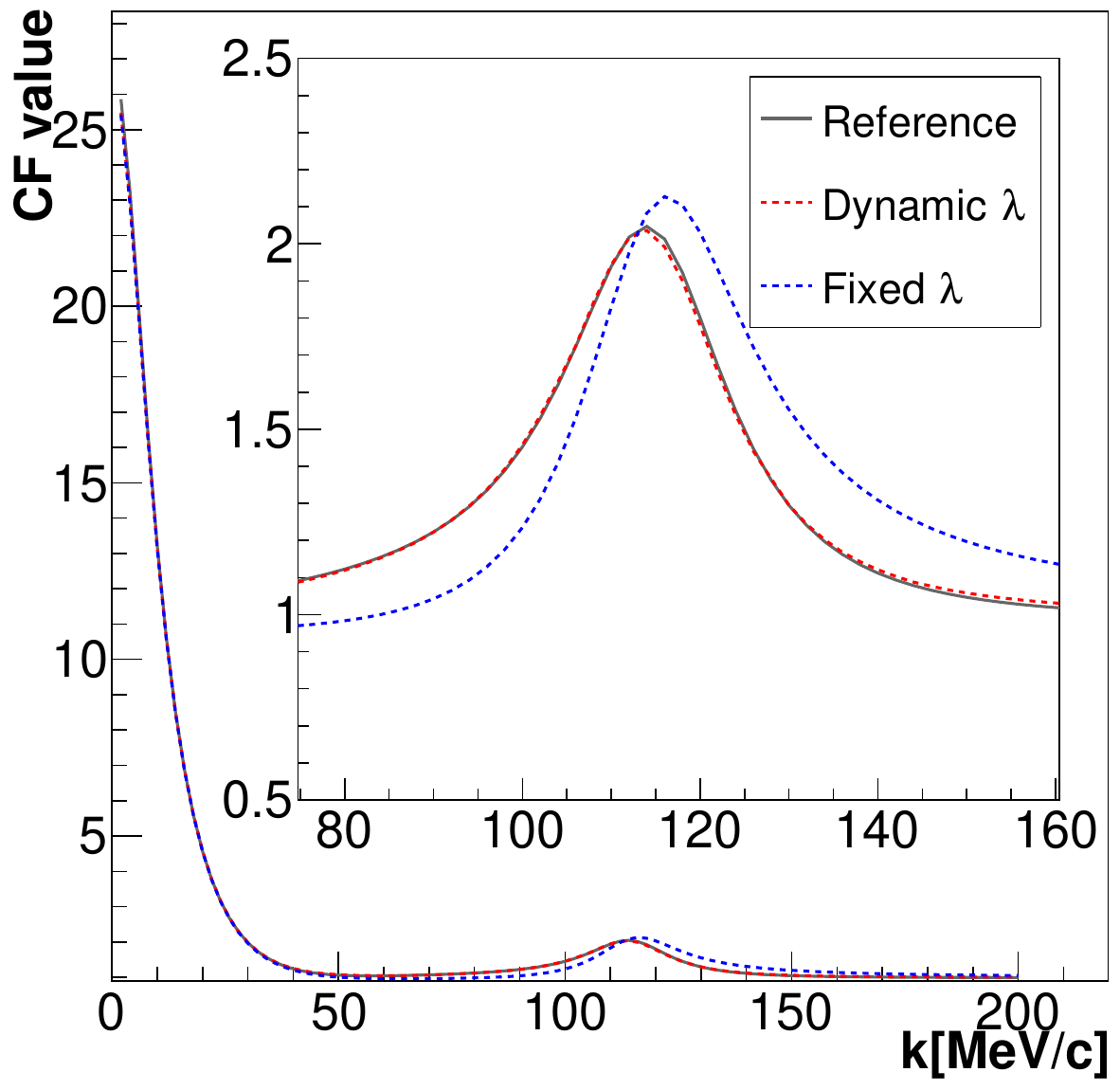}
    \caption{Modified neutron-neutron correlation functions in the spin-0 channel with a Gaussian source of $\sigma=4$\,fm. The reference curve is computed from the exact wave function. The curve labeled ``Dynamic $\lambda$'' is obtained by using variable values of $\lambda$ for each data point, where $\lambda$ is determined by minimizing the absolute value of the corresponding correction term over the range $\lambda = 0.3 - 4$\,fm. The curve labeled ``Fixed $\lambda$'' is obtained with $\lambda=1.3$\,fm for the s-wave and 1.8\,fm for the d-wave.}
    \label{fig:lambda_compare}
\end{figure}

We also employ dynamical values of $\lambda$ to make the correction terms vanish for each partial wave and each momentum. The improvement of the correlation function obtained by replacing the fixed $\lambda$ with the dynamical one is shown in Fig.~\ref{fig:lambda_compare}. The peak around $k=115$\,MeV/c in the correlation function arises from the enhancement of the d-wave interaction. It corresponds to a zero of Eq.\eqref{d-wave fit}, where $\delta_2$ passes through $\pi/2$ and the correction term vanishes. A static $\lambda$ cannot capture both sides of the peak because of the pronounced sign reversal of the correction term, so a dynamical $\lambda$ is necessary and proves effective.

The locations of the peaks in the correlation function also determine the number of coefficients required in Eqs.~\eqref{ERE_nC} and \eqref{ERE_C}. The s-wave interaction produces the first peak near $k=0$ in Fig.~\ref{fig:lambda_compare}, whereas the second peak, induced by the d-wave interaction, lies at the higher momentum $k=115$Mev/c and requires more terms in the Taylor expansion to reduce the remainder. We find that at least four terms are needed to describe the behavior of the d-wave peak, as given in Eq.~\eqref{d-wave fit}.

\begin{table}[htbp]
  \centering
  \caption{Interaction parameters extracted from the simulated correlation function.}
  \label{tab:table1}
  \begin{ruledtabular}
  \begin{tabular}{lccc}
    Parameter & Extracted value\footnotemark[1] & Reference\footnotemark[1] & Relative deviation\\
    \colrule
    $p_{0,0}$ & 0.1141 & 0.1155 & 1.2\%\\
    $p_{0,1}$ & 2.7909 & 2.7364 & 2.0\%\\
    
    $p_{2,0}$ & 0.1543 & 0.1414 & 9.1\%\\
    $p_{2,1}$ & 1.7723 & 1.7985 & 1.5\%\\
    $p_{2,2}$ & -5.1234 & -5.1068 & 0.3\%\\
    $p_{2,3}$ & -3.8291 & -3.8104 & 0.5\%\\
  \end{tabular}
  \footnotetext[1]{\raggedright Rescaled parameters in natural units\par}
  \end{ruledtabular}

\end{table}

The fitting result is shown in Table~\ref{tab:table1}, where the reference values are obtained from the variable phase method with ninth-order effective range expansion. The physical meanings of the coefficients are explained following Eq.~\eqref{ERE_nC}. Although the uncertainty of each point is not evaluated, the parameters are extracted by minimizing $\chi^2$ and deriving $\chi^2_{\mathrm{min}}\approx0.007$, with the uncertainty of every point treated uniformly as 1. 

The relative deviations for most parameters are small enough for practical applications. However, the first two parameters in Eq.~\eqref{d-wave fit} show larger deviations than others. Since the nontrivial d-wave contribution is localized around the peak at $k = 115$\,MeV/c, the parameters $p_{2,0}$ and $p_{2,1}$ that dominate at low momentum are only loosely constrained. As the partial wave order increases, we expect the fitting to focus on parameters of correspondingly higher order.

\section{Conclusions}

In this article, we have developed a femtoscopic approach to efficiently extract interaction parameters for all partial waves from the correlation function. All fitting parameters are defined within the effective range expansion. This model can also smoothly degenerate to the LL model by retaining only the s-wave and taking the limit $\lambda\to$ 0. Furthermore, we have successfully generated several correlation functions and investigated the dependence on $\lambda$ in the model, which is related to other models that employ cutoff wave functions. Finally, we have validated our method from multiple perspectives and applied it to simulated correlation data, obtaining results that closely agree with the reference values.

Our framework can be readily applied to realistic experimental data for spin-0 particle pairs, which evidently involve contributions from higher partial waves in the central-force case. For broader applicability, we plan to extend our model in the next paper to the noncentral-force case for spin-1/2 particle pairs, such as proton-neutron and proton-$\Lambda$.

\textbf{{Acknowledgement}} This work is supported by the National Natural Science Foundation of China under Grant Nos. 12335008 and 11927901, by the Ministry of Science and Technology under Grant No. 2022YFE0103400,  by the Center for High Performance Computing and Initiative Scientific Research Program in Tsinghua University.

\onecolumngrid
\appendix

\twocolumngrid
\bibliographystyle{apsrev4-2}  
\bibliography{apssamp}

@PREAMBLE{
 "\providecommand{\noopsort}[1]{}" 
 # "\providecommand{\singleletter}[1]{#1}%" 
}

@article{Recent1,
    author = "Epelbaum, Evgeny and Heihoff, Sven and Mei{\ss}ner, Ulf-G. and Tscherwon, Alexander",
    title = "{Can the Strong Interactions between Hadrons Be Determined Using Femtoscopy?}",
    eprint = "2504.08631",
    archivePrefix = "arXiv",
    primaryClass = "nucl-th",
    doi = "10.1103/tfsb-wlsd",
    journal = "Phys. Rev. Lett.",
    volume = "136",
    number = "21",
    pages = "212301",
    year = "2026"
}

@Article{Recent2,
author={Collaboration, The ALICE},
title={Observation of deuteron and antideuteron formation from resonance-decay nucleons},
journal={Nature},
year={2025},
month={Dec},
day={01},
volume={648},
number={8093},
pages={306-311},
issn={1476-4687},
doi={10.1038/s41586-025-09775-5},
url={https://doi.org/10.1038/s41586-025-09775-5}
}

@article{Recent3,
    author = "Si, Dawei and others",
    title = "{Extracting Neutron-Neutron Interaction Strength and Spatiotemporal Dynamics of Neutron Emission from the Two-Particle Correlation Function}",
    eprint = "2501.09576",
    archivePrefix = "arXiv",
    primaryClass = "nucl-ex",
    doi = "10.1103/PhysRevLett.134.222301",
    journal = "Phys. Rev. Lett.",
    volume = "134",
    number = "22",
    pages = "222301",
    year = "2025"
}

@article{specialbeam,
title = {Measurement of the nuclear multiplicity ratio for Ks0 hadronization at CLAS},
journal = {Physics Letters B},
volume = {706},
number = {1},
pages = {26-31},
year = {2011},
issn = {0370-2693},
doi = {https://doi.org/10.1016/j.physletb.2011.10.071},
url = {https://www.sciencedirect.com/science/article/pii/S0370269311013372},
author = {A. Daniel and others}
}

@article{Exotic1,
    author = "An, Jing",
    title = "{Measurements of proton{\textendash}{\ensuremath{\Xi}}{\ensuremath{-}} correlation functions in Au+Au collisions from the STAR beam energy scan II}",
    doi = "10.1016/j.jspc.2026.100501",
    journal = "J. Subatomic Part. Cosmol.",
    volume = "6",
    pages = "100501",
    year = "2026"
}

@article{Exotic2,
    author = "Takizawa, Makoto and Takeuchi, Sachiko and Yamaguchi, Yasuhiro and Hosaka, Atsushi",
    title = "{Femtoscopic study on \(D^{0}{\overline{D}}^{*0}\) interaction}",
    doi = "10.1016/j.jspc.2026.100389",
    journal = "J. Subatomic Part. Cosmol.",
    volume = "5",
    pages = "100389",
    year = "2026"
}

@article{Exotic3,
    author = "Zheng, Ru-You and Liu, Zhi-Wei and Geng, Li-Sheng",
    title = "{{\ensuremath{\Lambda}}cN correlation functions with leading-order covariant chiral interactions}",
    eprint = "2604.01154",
    archivePrefix = "arXiv",
    primaryClass = "hep-ph",
    doi = "10.1103/l9zj-f6c5",
    journal = "Phys. Rev. D",
    volume = "114",
    number = "1",
    pages = "014006",
    year = "2026"
}

@article{Exotic4,
  title = {Charmonium-nucleon femtoscopic correlation function},
  author = {Liu, Zhi-Wei and Ge, Duo-Lun and Lu, Jun-Xu and Liu, Ming-Zhu and Geng, Li-Sheng},
  journal = {Phys. Rev. D},
  volume = {112},
  issue = {5},
  pages = {054019},
  numpages = {8},
  year = {2025},
  month = {Sep},
  publisher = {American Physical Society},
  doi = {10.1103/3bdh-blwh},
  url = {https://link.aps.org/doi/10.1103/3bdh-blwh}
}

@article{Exotic5,
  title = {Study of the $DK$ interaction with femtoscopic correlation functions},
  author = {Liu, Zhi-Wei and Lu, Jun-Xu and Geng, Li-Sheng},
  journal = {Phys. Rev. D},
  volume = {107},
  issue = {7},
  pages = {074019},
  numpages = {8},
  year = {2023},
  month = {Apr},
  publisher = {American Physical Society},
  doi = {10.1103/PhysRevD.107.074019},
  url = {https://link.aps.org/doi/10.1103/PhysRevD.107.074019}
}

@ARTICLE{partialwavemethod,
       author = {{Fax{\'e}n}, H. and {Holtsmark}, J.},
        title = "{Beitrag zur Theorie des Durchganges langsamer Elektronen durch Gase}",
      journal = {Zeitschrift fur Physik},
         year = 1927,
        month = may,
       volume = {45},
       number = {5-6},
        pages = {307-324},
          doi = {10.1007/BF01343053},
       adsurl = {https://ui.adsabs.harvard.edu/abs/1927ZPhy...45..307F}
}

@ARTICLE{LL1,
author = {{Lednick{\'y}}, R. and {Lyuboshitz}, V.~L.},
        title = "{The influence of final state interaction on two-particle correlations in multiple production of particles and resonances}",
      journal = {Acta Physica Hungarica Heavy Ion Physics},
         year = 1996,
        month = apr,
       volume = {3},
       number = {1-2},
        pages = {93-113},
          doi = {10.1007/BF03053635},
       adsurl = {https://ui.adsabs.harvard.edu/abs/1996APHHI...3...93L}
}

@techreport{LL2,
author        = "Lednicky, R and Lyuboshitz, V L",
      title         = "{Influence of final-state interaction on correlations of
                       two particles with nearly equal momenta}",
      institution   = "Joint Inst. Nucl. Res.",
      reportNumber  = "JINR-E2-81-453, JINR-P2-81-453",
      address       = "Dubna",
      year          = "1981",
      url           = "https://cds.cern.ch/record/131203",
}

@article{LLapp1,
    author = "Pradhan, Raghunath",
    collaboration = "CMS",
    title = "{Results of femtoscopic correlations at CMS}",
    doi = "10.1016/j.jspc.2025.100144",
    journal = "J. Subatomic Part. Cosmol.",
    volume = "4",
    pages = "100144",
    year = "2025"
}

@article{LLapp2,
    author = "Mikhaylov, Konstantin",
    editor = "Teterin, Petr",
    collaboration = "ALICE",
    title = "{Non-identical charged kaon femtoscopy in Pb{\textemdash}Pb collisions at $\sqrt{s_{NN}}$ = 2,76 TeV by ALICE}",
    doi = "10.1088/1742-6596/1690/1/012099",
    journal = "J. Phys. Conf. Ser.",
    volume = "1690",
    number = "1",
    pages = "012099",
    year = "2020"
}

@article{HigherWavePrevious1,
title = {Higher partial waves in femtoscopy},
journal = {Journal of Subatomic Particles and Cosmology},
volume = {3},
pages = {100017},
year = {2025},
issn = {3050-4805},
doi = {https://doi.org/10.1016/j.jspc.2024.100017},
url = {https://www.sciencedirect.com/science/article/pii/S3050480524000177},
author = {Koichi Murase and Tetsuo Hyodo}
}

@article{HigherWavePrevious2,
    author = "Liu, Si-Wei and Xie, Ju-Jun",
    title = "{Femtoscopic Correlation Functions for General Partial Waves: Application to the {\ensuremath{\Lambda}}(1520) Resonance}",
    eprint = "2601.22695",
    archivePrefix = "arXiv",
    primaryClass = "hep-ph",
    doi = "10.1088/0256-307X/43/5/050202",
    journal = "Chin. Phys. Lett.",
    volume = "43",
    number = "5",
    pages = "050202",
    year = "2026"
}

@article{widerangeofsys1,
    author = "Acharya, Shreyasi and others",
    collaboration = "ALICE",
    title = "{Space-time evolution of particle emission in p{\textendash}Pb collisions at $\sqrt{{s}_{\text{NN}}}=5.02$ TeV with 3D kaon femtoscopy}",
    eprint = "2601.15081",
    archivePrefix = "arXiv",
    primaryClass = "nucl-ex",
    reportNumber = "CERN-EP-2025-237",
    doi = "10.1007/JHEP04(2026)113",
    journal = "JHEP",
    volume = "04",
    pages = "113",
    year = "2026"
}

@phdthesis{widerangeofsys2,
    author = "Heybeck, Benedict",
    title = "{Measurement of the $\Sigma^+$ production and the p-$\Sigma^+$ correlation function in proton-proton collisions with ALICE at the LHC}",
    doi = "10.21248/gups.94367",
    school = "Goethe U., Frankfurt (main)",
    year = "2025"
}

@article{HigherWaveExperi1,
    author = "Acharya, Shreyasi and others",
    collaboration = "ALICE",
    title = "{Accessing the strong interaction between {\ensuremath{\Lambda}} baryons and charged kaons with the femtoscopy technique at the LHC}",
    eprint = "2305.19093",
    archivePrefix = "arXiv",
    primaryClass = "nucl-ex",
    reportNumber = "CERN-EP-2023-106",
    doi = "10.1016/j.physletb.2023.138145",
    journal = "Phys. Lett. B",
    volume = "845",
    pages = "138145",
    year = "2023"
}

@article{HigherWaveExperi2,
    author = "Acharya, Shreyasi and others",
    collaboration = "ALICE",
    title = "{Scattering studies with low-energy kaon-proton femtoscopy in proton-proton collisions at the LHC}",
    eprint = "1905.13470",
    archivePrefix = "arXiv",
    primaryClass = "nucl-ex",
    reportNumber = "CERN-EP-2019-118",
    doi = "10.1103/PhysRevLett.124.092301",
    journal = "Phys. Rev. Lett.",
    volume = "124",
    number = "9",
    pages = "092301",
    year = "2020"
}

@article{HigherWaveExperi3,
    author = "Acharya, Shreyasi and others",
    collaboration = "ALICE",
    title = "{Kaon{\textendash}proton strong interaction at low relative momentum via femtoscopy in Pb{\textendash}Pb collisions at the LHC}",
    eprint = "2105.05683",
    archivePrefix = "arXiv",
    primaryClass = "nucl-ex",
    reportNumber = "CERN-EP-2021-080",
    doi = "10.1016/j.physletb.2021.136708",
    journal = "Phys. Lett. B",
    volume = "822",
    pages = "136708",
    year = "2021"
}

@article{ElecQuad1,
  title = {An Electrical Quadrupole Moment of the Deuteron The Radiofrequency Spectra of HD and ${\mathrm{D}}_{2}$ Molecules in a Magnetic Field},
  author = {Kellogg, J. M. B. and Rabi, I. I. and Ramsey, N. F. and Zacharias, J. R.},
  journal = {Phys. Rev.},
  volume = {57},
  issue = {8},
  pages = {677--695},
  numpages = {0},
  year = {1940},
  month = {Apr},
  publisher = {American Physical Society},
  doi = {10.1103/PhysRev.57.677},
  url = {https://link.aps.org/doi/10.1103/PhysRev.57.677}
}

@article{ElecQuad2,
title = {A study of deuteron quadrupole moment effects in sub-coulomb scattering of tensor polarized deuterons},
journal = {Nuclear Physics A},
volume = {435},
number = {2},
pages = {502-522},
year = {1985},
issn = {0375-9474},
doi = {https://doi.org/10.1016/0375-9474(85)90475-0},
url = {https://www.sciencedirect.com/science/article/pii/0375947485904750},
author = {J.E. Kammeraad and L.D. Knutson}
}

@ARTICLE{AHP,
title = {Final state interactions in two-particle interferometry},
  author = {Anchishkin, D. and Heinz, U. and Renk, P.},
  journal = {Phys. Rev. C},
  volume = {57},
  issue = {3},
  pages = {1428--1439},
  numpages = {0},
  year = {1998},
  month = {Mar},
  publisher = {American Physical Society},
  doi = {10.1103/PhysRevC.57.1428},
  url = {https://link.aps.org/doi/10.1103/PhysRevC.57.1428}
}

@article{KooninPratt,
title = {Proton pictures of high-energy nuclear collisions},
journal = {Physics Letters B},
volume = {70},
number = {1},
pages = {43-47},
year = {1977},
issn = {0370-2693},
doi = {https://doi.org/10.1016/0370-2693(77)90340-9},
url = {https://www.sciencedirect.com/science/article/pii/0370269377903409},
author = {Steven E. Koonin}
}

@incollection{QM1,
title = {CHAPTER XVII - ELASTIC COLLISIONS},
editor = {L.D. LANDAU and E.M. LIFSHITZ},
booktitle = {Quantum Mechanics (Third Edition)},
publisher = {Pergamon},
edition = {Third Edition},
pages = {502-590},
year = {1977},
isbn = {978-0-08-020940-1},
doi = {https://doi.org/10.1016/B978-0-08-020940-1.50024-4},
url = {https://www.sciencedirect.com/science/article/pii/B9780080209401500244},
author = {L.D. LANDAU and E.M. LIFSHITZ}
}

@misc{Math1,
key = "{\relax DLMF}",
       title = "{\it NIST Digital Library of Mathematical Functions}",
howpublished = "\url{https://dlmf.nist.gov/}, Release 1.2.4 of 2025-03-15",
         url = "https://dlmf.nist.gov/",
        note = "F.~W.~J. Olver, A.~B. {Olde Daalhuis}, D.~W. Lozier, B.~I. Schneider,
                R.~F. Boisvert, C.~W. Clark, B.~R. Miller, B.~V. Saunders,
                H.~S. Cohl, and M.~A. McClain, eds."
}

@ARTICLE{EffRange_nc,
 title = {Effective-range function methods for charged particle collisions},
  author = {Gaspard, David and Sparenberg, Jean-Marc},
  journal = {Phys. Rev. C},
  volume = {97},
  issue = {4},
  pages = {044003},
  numpages = {15},
  year = {2018},
  month = {Apr},
  publisher = {American Physical Society},
  doi = {10.1103/PhysRevC.97.044003},
  url = {https://link.aps.org/doi/10.1103/PhysRevC.97.044003}
}

@ARTICLE{QM2,
title={Quantum Mechanics, 3rd ed.},
  author={Leonard I. Schiff and John H. Gardner},
  journal={American Journal of Physics},
  year={1973},
  volume={41},
  pages={599-600},
  url={https://api.semanticscholar.org/CorpusID:120668893}
}

@ARTICLE{Math2,
  title={Handbook of Mathematical Functions With Formulas, Graphs and Mathematical Tables (National Bureau of Standards Applied Mathematics Series No. 55)},
  author={Milton Abramowitz and Irene A. Stegun and David Miller},
  journal={Journal of Applied Mechanics},
  year={1965},
  volume={32},
  pages={239-239},
  url={https://api.semanticscholar.org/CorpusID:121782574}
}

@article{ERE_C,
  title = {Formulas for the $\ensuremath{\delta}$-shell-plus-Coulomb potential for all partial waves},
  author = {Kok, L. P. and de Maag, J. W. and Brouwer, H. H. and van Haeringen, H.},
  journal = {Phys. Rev. C},
  volume = {26},
  issue = {6},
  pages = {2381--2396},
  numpages = {0},
  year = {1982},
  month = {Dec},
  publisher = {American Physical Society},
  doi = {10.1103/PhysRevC.26.2381},
  url = {https://link.aps.org/doi/10.1103/PhysRevC.26.2381}
}

@article{ReidPotential,
title = {Local phenomenological nucleon-nucleon potentials},
journal = {Annals of Physics},
volume = {50},
number = {3},
pages = {411-448},
year = {1968},
issn = {0003-4916},
doi = {https://doi.org/10.1016/0003-4916(68)90126-7},
url = {https://www.sciencedirect.com/science/article/pii/0003491668901267},
author = {Roderick V Reid}
}

@article{VPEextension,
  title = {Extended variable phase method for spin-1/2 correlation functions},
  author = {Zou, Renjie and Xiao, Sheng and Qin, Zhi and Xiao, Zhigang},
  journal = {Phys. Rev. C},
  volume = {114},
  issue = {1},
  pages = {014003},
  numpages = {10},
  year = {2026},
  month = {Jul},
  publisher = {American Physical Society},
  doi = {10.1103/kxrh-dmyy},
  url = {https://link.aps.org/doi/10.1103/kxrh-dmyy}
}

\end{document}